\documentclass[11pt]{article}

\usepackage{graphicx}

\begin{document}

\vskip 2cm
\title{Quantum secret sharing based on reusable GHZ states as
secure carriers}
\author{Saber Bagherinezhad $^{a}$
\footnote{e-mail:bagherin@ce.sharif.ac.ir}\\
 Vahid Karimipour $^b$\footnote{e-mail:
 vahid@sina.sharif.ac.ir.}\\
 $^a$Department of Computer Science, Sharif University of Technology,\\
 $^b$Department of Physics, Sharif University of Technology\\
 P. O. Box 11365-9161, Tehran, Iran }
\maketitle

\begin{abstract}
 We introduce a protocol for quantum secret sharing based on
 reusable entangled states. The entangled state between the sender
 and the receiver acts only as a carrier to which data bits are entangled by the sender and disentangled from it by
 the receivers, all by local actions of simple gates.
 We also show that the interception by Eve or the cheating of one of the receivers introduces a quantum bit error rate (QBER) larger than
 25 percent which can be detected by comparing a subsequence of the bits.
\end{abstract}
\newpage

\section{Introduction}
 The past few years has witnessed progress in theoretical aspects and
 experimental implementations of quantum cryptography.
(For an elementary introduction to the subject
see \cite{vol} and for a comprehensive review of recent advances see \cite{rev}.)\\
 One of the desirable protocols for secure communication is called
 secret sharing, the simplest of which is when the sender Alice wants to send a secret message to two
 receivers Bob and Charlie so that non of the receivers can recover the message on
 his own.
  In 1998, Hillery, Buzek, and Berthiaume proposed
 a quantum solution for secret sharing \cite{buzek}. In their method which is
 inspired by the quantum key distribution method of Bennett and
 Brasard \cite{bb84}, and Ekert\cite{ekert}, Alice, Bob and
 Charlie share a Greenberger-Horne-Zeilinger (GHZ) state
 \cite{ghz}. They then carry out measurements of their bits in
 either of the two non-commuting bases, say $x$ and $y$ bases in random. Since the results or measurements
are correlated for half of the cases, they can establish a secret
key between themselves by announcing their bases of measurements.
 The
aim of the present article which is inspired by the work of
Zhang, Li and Guo \cite{zlg} in
 quantum key distribution, is to present an alternative method for
 secret sharing, which is based on sharing entangled states as carriers to which data bits are entangled by the sender Alice and
 disentangled by the receivers Bob and Charlie. The role of this carrier is to make communication secure against
 intervention of evesdroppers or cheating of any of the
 receivers.\\
Two remarks are in order comparing the differences of our protocols with the known ones \cite{buzek,carlsson}.\\
1- If
 experimental realization of constructing and distributing $GHZ$
 states to parties at long separation becomes a possibility in the future, it is
 plausible to assume that the maintenance of the correlations of these states
 will be easier and less costly than their creation anew for
 every round as is needed in earlier protocols for secret
 sharing.\\
2- We should also point out that this protocol is not far from
the reach of the near future experiments. At present single qubit
gates and double qubit gates very close to the CNOT gate, can be
implemented on individual ions where up to ten ions are kept in a
coherent state \cite{blatt}. One can also imagine that by methods
similar to the ones proposed in \cite{cabrillo}, distant atoms or
ions can be entangled with each other in the near future. The
structure
 of this paper is as follows: In section 2 we introduce the
 protocol for two receiving parties, and discuss how information is split and is protected from unauthorized parties.
 We also show how intervention of Eve can be detected. We end up the paper with conclusions.
\section{The secret sharing protocol with two parties}
 Suppose Alice wants to sent a message which is already in the
 form of a sequence of classical bits $ q_1, q_2, q_3, \cdots $,
 ($q_i = 0\ \  {\rm {or}}\ \ 1$) to Bob and Charlie, so that the
 receivers can infer this message only by their mutual assistance.
 We begin with our notations.
 We use subscripts $a$, $b$, $c$ and $e$ on states
 and operators for Alice, Bob, Charlie, and Eve respectively. Any
 other space carrying message qubits is specified by numerical
 subscripts, 1, 2, etc. A controlled gate like CNOT is denoted simply by $C$, and is
 specified by two subscripts, the first one is the control bit,
 the second is the target bit. Thus $C_{a1}$ is the controlled NOT
 gate which is controlled by Alice qubit and acts on the qubit in
 line 1, i.e: $
 C_{a1}|q,q'\rangle_{a1} = |q, q+q'\rangle_{a1}
$,
 where $ q $ and $q' $ are $ 0 $ and $1 $ and addition is
 performed mod two.\\
 The Hadamard gate acts as $ H|0\rangle = \frac{1}{\sqrt{2}}(|0\rangle + |1\rangle)\ $  and
 $\ H|1\rangle = \frac{1}{\sqrt{2}}(|0\rangle - |1\rangle)
$.
 By $|\overline{0}\rangle $ and  $ |\overline{1}\rangle $, we mean two qubit states which are uniform
 superposition of basis states the sum of whose digits modulo $2$
 are respectively $ 0 $ and $ 1$, i.e: $
   |\overline {0}\rangle = \frac{1}{\sqrt{2}}\big(|0,0\rangle + |1,
   1\rangle\big) $ and $ |\overline {1}\rangle = \frac{1}{\sqrt{2}}\big(|0,1\rangle + |1,
   0\rangle\big) $, which can be identified only by the
 collaboration of the two parties possessing the bits.\\
Moreover the following property is also easily verified:
\begin{equation}\label{chi2}
C_{a1}C_{b2}|\overline{q}\rangle_{a,b}|\overline{q'}\rangle_{1,2}= |\overline{q}\rangle_{a,b}|\overline{q+q'}\rangle_{1,2}.
\end{equation}
  We also need to define two three-particle states, namely the GHZ state which we denote by
 \begin{equation}\label{g3}
 |G\rangle := \frac{1}{\sqrt{2}}(|000\rangle
 + |111\rangle).
 \end{equation}
 and an even parity state which we denote by
 \begin{equation}\label{e3}
 |E\rangle := \frac{1}{2}(|000\rangle+ |110\rangle
 + |101\rangle + |011\rangle)\equiv \frac{1}{\sqrt{2}}(|0\rangle|\overline{0}\rangle+ |1\rangle|\overline{1}\rangle).
 \end{equation}
 These two states are transformed to each other by the local operation of Hadamard gates, that is:
 \begin{equation}\label{g3e3}
 |G\rangle = H\otimes H \otimes H |E\rangle \hskip 1cm {\rm and} \hskip 1cm |E\rangle
 = H\otimes H \otimes H |G\rangle \end{equation}
 We will use these two states which are shared by all three parties, as carriers of information,
 (the $|G\rangle$ state in the odd rounds and the $|E\rangle$ state in the even rounds).
Alice entangles her data bits to the above carriers and bob and
Charlie disentangle the data bits from these carriers. Due to
property (\ref{g3e3}), the action of Hadamard gates performed by
all the parties at the end of each round switches the Carrier to
the appropriate one for the next round. This switching of
carriers is also crucial for the security of the protocol as we
will see in the sequel.\\
 For sending a classical bit $q$ Alice
 may encode it as a state $|q,q\rangle$ and send this state simply to Bob and Charlie. At the
 destination Bob and Charlie can measure their
 corresponding bits and recover the bit $q$. In this case Bob and
 Charlie can understand $q$ without each other's assistance, and
 in fact Alice sends half of the bits say the odd numbered ones
 $q_1, q_3, q_5 \cdots $ in this way. For the other half, $ q_2,
 q_4, q_6 \cdots $, she encodes a bit $q$ in the form of a state $ |\overline{q}\rangle =
 \frac{1}{\sqrt{2}}\big(|0,q\rangle + |1,q+1\rangle\big)$ and
 sends it to Bob and Charlie who are assumed to have access to the
first and second spaces respectively. Any such state gives no information at all to either Bob
 or Charlie, since the density matrix of each of them is easily
 seen to be completely mixed. However they can identify the bit
 $q$ by communicating to each other the result of their
 measurements. The value of the bit $q$ is simply obtained by
 adding their result mod 2. In this way Alice can split a message
 so that Bob and Charlie can recover the message only by their
 cooperation. A cheating of the kind of wrong declaration of the
 results by one of the receivers leads to 50 percent errors which
 is easily detected by comparing a subsequence of the bits
 received with those actually sent by Alice. By this comparison
 Alice finds that at least one of the receivers has been dishonest.
 Although she can not determine which one.\\
 This is the part of protocol which deals with splitting of
 information. Now we are faced with the problem of protecting
 information against Eavesdropping and against cheating of one of
 the parties who may find access via the collaboration of Eve
 to both the qubits.\\
 We should assume that the quantum channel used by Alice for
 sending the qubits is not secure and can be penetrated by an
 unauthorized third party called Eve (who incidentally may be one of the dishonest receivers say Bob)
 finding access to both of the
bits in transition and retrieving the data (without assistance of
Charlie). We now make our protocol safe against such attacks or
cheating.
 The strategy is to entangle the message qubits with an already
 entangled state in possession of Alice, Bob and Charlie, in a
 highly mixed form, so that while being sent, these qubits if
 accessed by Eve or by one of the receivers say Bob, carry no information at all.
 Moreover we should also show that Eve's intervention
and Bob's cheating can be detected by the other parties. We use two
 different forms of carriers for odd and even bits. For odd
 bits, we proceed as follows. Alice entangles the state $
 |qq\rangle_{12}$ to the already present GHZ state
 $|G\rangle_{abc}$ by performing CNOT gates $C_{a1}C_{a2}$ on
\begin{equation}
|G\rangle_{abc}|qq\rangle_{12} =
\frac{1}{\sqrt{2}}\big(|0,0,0\rangle
 + |1,1,1\rangle)_{a,b,c} |q,q\rangle_{12}
\end{equation}
 to produce the state:
 \begin{eqnarray}\label{phi0}
 |\Phi^{odd}\rangle &=& \frac{1}{\sqrt{2}}\Big(|0,0,0\rangle_{abc}
 |{q,q}\rangle_{12} + |1,1,1\rangle_{a,b,c} |1+q,1+q\rangle_{12}\Big)
 \end{eqnarray}
 At the destination, Bob and Charlie act on this
 state by the operators $ C_{b1} $ and $C_{c2}$ and extract the
 state $ |q,q\rangle_{1,2} $ where each one of them can
 read independently his own bit. By her action Alice has entangled
 the double bit $ |q,q\rangle _{1,2}$ so that while in
 transmission it is a mixture of  $ |q,q\rangle $ and
 $ |1+q,1+q\rangle $ which conveys no information to Eve about the value of the bit
 being sent.\\
It is also seen from the state (\ref{phi0}) that a simple
intercept-resend strategy adopted by Eve of the two flying data
qubits, will make $50$ percent error in the data bits jointly
received by
Bob and Charlie with those sent by Alice. Therefore Eve's presence can be detected by publicly comparing a subsequence of the bits sent by Alice
with those received by Bob and Charlie.\\
 For the even bits which are encoded as states $ |\overline{q}\rangle $,
 i.e. ($ |\bar{0}\rangle= \frac{1}{\sqrt{2}}(|00\rangle + |11\rangle) $
 and $ |\bar{1}\rangle= \frac{1}{\sqrt{2}}(|01\rangle + |10\rangle)$),
 Alice entangles this state to the carrier $ |E\rangle $ by performing only one single CNOT gate
 $C_{a1}$ on
\begin{equation}
 |E\rangle_{abc}\otimes
 |\overline{q} \rangle_{12} = \frac{1}{\sqrt{2}}(|0\rangle|\overline{0}\rangle+ |1\rangle|\overline{1}\rangle)_{abc}|\overline{q}\rangle_{1,2}
 \end{equation}
to produce the state
 \begin{equation}\label{psi1}
 |\Psi^{even}\rangle =  \frac{1}{\sqrt{2}}\Big(|0\rangle_a |\overline
 {0}\rangle_{bc} |\overline {q}\rangle_{12}+ |1\rangle_a
 |\overline {1}\rangle_{bc} |\overline {1+q}\rangle_{12}\Big)
 \end{equation}
 where we have used the the property (\ref{chi2}).
 At the
 destination, Bob and Charlie act on this state by the operators $
 C_{b1} $ and $C_{c2}$ where again by (\ref {chi2}) they extract $
 |\overline {q}\rangle_{1,2}$ which they can identify
 completely only by their collaboration. It is quite simple to see from (\ref{psi1}) that
   $\rho_{b1} = \rho_{c2} = \rho_{12} =\frac{1}{2}I $.

This means that neither Eve alone who may supposedly find access
to the two data bits nor any of the receivers independently, can
find the data bit which has been encoded and sent by Alice. For
the even rounds the simple intercept-resend strategy of Eve
introduces $50$ percent discrepancy among the data bits of Bob
with those of Charlie which again leads to the detection of Eve,
since a state which has been encoded as $ |\bar{0}\rangle =
\frac{1}{\sqrt{2}}(|00\rangle +|11\rangle) $ is received half the
time by Bob
and Charlie as a state $ |\bar{1}\rangle = \frac{1}{\sqrt{2}}(|01\rangle +|10\rangle)$. \\
Note that once the data bits are measured by Bob and Charlie,
only one of them needs to publicly announce the result of his
measurement, and the other will find the actual bit sent by
Alice, by simply adding the bit publicly announced to the one that
he has actually measured. This public announcement again does not
convey any information to Eve. Moreover a wrong declaration of
results by one of the receivers
say Bob again leads to discrepancies of the bits between Alice and Charlie.\\
Before going on to study a general attack of Eve or Cheating of Bob, lets finish the
protocol by saying how
 Alice, Bob and Charlie switch their entangled state from the GHZ
 state $ |G\rangle_{a,b,c}$ for odd rounds to the even state $|E\rangle _{a,b,c}$ state for even rounds. They can do
 this simply by performing Hadamard gates on their respective
 states at the end of every round of the protocol. The reason is
 relation (\ref{chi2}).
 Thus they start the first round with
 the GHZ state $|G\rangle$ and end up with the $| E\rangle$ state which is used for the
 second round. At the end of the second round they have produced
 again the GHZ state $|G\rangle $ which will be used for
 the next round and so on.\\
We now assume that Eve who may be a collaborator of one of the
receivers say Bob follows a more complicated strategy by
entangling her system (ancilla)
 with the states of Alice, Bob and Charlie in the most general form, that is:
 \begin{equation}\label{detection}
   |\Theta\rangle_{a,b,c,e} = \sum_{i,j,k}
   |i,j,k\rangle_{abc}\otimes \eta_{ijk} \hskip 2cm i,j,k = 0,1.
 \end{equation}
 where $\eta_{i,j,k} $'s are un-normalized states of Eve. Eve wants to make this entanglement so that
 at the end of each round of sending and receiving a bit, useful
 information about that bit is collected in her ancilla which she
 can measure safely later on. Consider an odd round of the process
 and suppose that the starting state of Alice, Bob and Charlie,
 ignorant of the presence of Eve is as above. Eve is clever enough
 to entangle her state such that she does not perturb the values
 of the final bits measured by Bob and Charlie when the protocol is
 run for this round. (otherwise a comparison of a substring of bits between Alice and Bob and Charlie, will reveal her presence
 or her collaboration with one of the receivers.).  She then finds that the ideal form of
 entanglement is as follows:
 \begin{equation}\label{detection2}
   |\Theta^{odd}\rangle =
   |0,0,0\rangle \eta_{000}+ |1,1,1\rangle\eta_{111}.
 \end{equation}
 where we have suppressed the subscripts on the states and the
 $\otimes $ symbol.
 If she keeps any other state in
 (\ref{detection2}), say a state like $|0,1,1\rangle \eta_{011}$,
 the total superposition
 will have a term $|0,1,1, q, q\rangle \eta_{011}$ which reveals
 the encoded bit $q$ by Alice as $ 1+q$ to Bob and Charlie. Or if
 she keeps a state like $|0,1,0\rangle \eta_{010}$,
the total superposition
 will have a term $|0,1,0, q, q\rangle \eta_{011}$ which reveals
 the encoded bit $q$ by Alice as $ 1+q$ to Bob and as $q$ to Charlie.
 In all cases these lead to her detection after a subsequence of bits are compared.
 A similar analysis reveals
 to Eve that the ideal form of entanglement for an even round of
 the protocol is:
 \begin{equation}\label{detection3}
   |\Theta^{even}\rangle =
   |0,0,0\rangle\xi_{000}+
   |1,1,0\rangle\xi_{110}+
   |1,0,1\rangle\xi_{101}+
   |0,1,1\rangle\xi_{011}
 \end{equation}
 The crucial point is that the Hadamard gates at the end of each
 round do not allow Eve to have desirable entanglement for every
 round. Eve can have desirable entanglement only if $ H^{\otimes
 3} |\Theta^{even}\rangle = |\Theta^{odd}\rangle $ and vise versa. A simple calculation yields:

  \begin{eqnarray}\label{detection5}
   H^{\otimes 3}|\Theta^{odd}\rangle = &\frac{1}{2\sqrt{2}}&(|0,0,0\rangle + |110\rangle +
   |1,0,1\rangle + |011\rangle )(\eta_{000}+
 \eta_{111})\cr  + &\frac{1}{2\sqrt{2}}&(|1,1,1\rangle + |0,0,1\rangle + |010\rangle +
 |100\rangle)(\eta_{000}- \eta_{111}).
 \end{eqnarray}
Equating this to $|\Theta^{even}\rangle $ yeilds $
 \eta_{000}=\eta_{111} \ $ and $\   \xi_{000}= \xi_{110} = \xi_{101} =
 \xi_{011} $.
Looking back at (\ref{detection2}) and (\ref{detection3}), we see
that this implies
 that this switching between desirable
 entanglement at alternative rounds is possible for Eve only if
 there is no entanglement at all in any of the rounds!
One may argue that Eve may not want to completely avoid any error
introduced into the data and she may entangle her system to the
carriers in order to reduce the quantum bit error rate (QBER) as
low as possible, lower than the expected level of noise and hence
escape the detection . We will show in the appendix, that Eve can
not lower the QBER averaged over odd and even rounds, below 25
percent. In this way we have shown the security of the protocol
against Eve's attack or Bob's Cheating.\\
In conclusion we have presented a new protocol for quantum secret
sharing based on reusable entangled states. In
 our protocol a sequence of bits is transmitted to two parties so
 that they can recover half of the bits independently and for the
 rest half they need to collaborate to find the identity of the
 bits. The distinctive feature of this protocol is the existence of a carrier which
carries the data bits from the sender to the receiver in secure
form, without any need for measurements in random bases and
public announcements. We have also discussed the security of the
protocol against evesdropping and against cheating of one of the
receivers,
 and have shown that any such action leads to high error rates in the sequence of bits between the participants.
 A rigorous proof of the security of the protocol is however beyond the scope of the present work.

\section{Appendix}
This appendix is a completion of section 3, where we show that
even if we Eve allows a small quantum bit error rate introduced
into the data, she can not find any form of entanglement of her
system to the carriers to achieve this goal. Suppose that in the
odd and even rounds Eve entangles her system to the carriers in
the following general forms:
\begin{equation}
 |\Theta^{odd}\rangle = |0\rangle \eta_{0}+ |1\rangle \eta_{1}+ |2\rangle \eta_{2}+|3\rangle \eta_{3}
+|4\rangle \eta_{4}+
  |5\rangle \eta_{5}+|6\rangle \eta_{6}+|7\rangle \eta_{7}.
\end{equation}
and
\begin{equation}
 |\Theta^{even}\rangle = |0\rangle \xi_{0}+|1\rangle \xi_{1}+ |2\rangle \xi_{2}+|3\rangle \xi_{3}
+|4\rangle \xi_{4}+
  |5\rangle \xi_{5}+|6\rangle \xi_{6}+|7\rangle \xi_{7}.
\end{equation}
where for simplicity we have used binary notation and suppressed
indices, i.e. $ |0\rangle \xi_0 \equiv |000\rangle_{abc}\otimes
\xi_0 , |3\rangle \xi_3 \equiv |011\rangle_{abc}\otimes \xi_3 $
etc. In order to reduce the QBER (the probability of unwanted bits
introduced into the transferred bits) below a tolerable threshold
$\epsilon$, she should choose the states such that:

\begin{equation}\label{1}
|\eta_1|^2 + |\eta_2|^2 +|\eta_3|^2 +|\eta_4|^2 +|\eta_5|^2
+|\eta_6|^2 \leq \epsilon,\hskip 1cm or \hskip 1cm  |\eta_0|^2 +
|\eta_7|^2 \geq 1-\epsilon
\end{equation}
in the odd rounds and
\begin{equation}\label{2}
|\xi_1|^2 + |\xi_2|^2 +|\xi_4|^2 +|\xi_7|^2 \leq \epsilon .
\end{equation}
However the action of Hadamard gates relates the two states
$|\Theta^{odd}\rangle$ and $|\Theta^{even}\rangle$ and hence the
states $ \{\eta_i\} $ and $\{\xi_i\}$. It is easy to find that
under the Hadamard operations:

\begin{eqnarray}\label{3}
\xi_1 &=& \frac{1}{2\sqrt 2}(\eta_0 + \eta_1 -\eta_2 -\eta_3
-\eta_4 -\eta_5 +\eta_6 +\eta_7) \cr \xi_2 &=& \frac{1}{2\sqrt
2}(\eta_0 - \eta_1 +\eta_2 -\eta_3 -\eta_4 +\eta_5 -\eta_6
+\eta_7) \cr \xi_4 &=& \frac{1}{2\sqrt 2}(\eta_0 + \eta_1 +\eta_2
+\eta_3 -\eta_4 -\eta_5 -\eta_6 -\eta_7) \cr \xi_7 &=&
\frac{1}{2\sqrt 2}(\eta_0 - \eta_1 -\eta_2 +\eta_3 -\eta_4 +\eta_5
+\eta_6 -\eta_7)
\end{eqnarray}
A simple rearrangement yields
\begin{eqnarray}\label{4}
 & & |\xi_1|^2 + |\xi_2|^2 + |\xi_4|^2 + |\xi_7|^2 = \cr
& &\frac{1}{2}( |\eta_0-\eta_4|^2 +
|\eta_1-\eta_5|^2+|\eta_2-\eta_6|^2+|\eta_3-\eta_7|^2)\cr & &\geq
\frac{1}{2}( |\eta_0-\eta_4|^2 + |\eta_3-\eta_7|^2) \simeq
\frac{1-\epsilon}{2}
\end{eqnarray}
where in the last line we have used (\ref{1}). Therefore we see
that keeping the QBER below a very small threshold $\epsilon$ in
the odd rounds, will introduce a QBER of about 50 percent in the
even rounds and vice versa. The average QBER over the odd and
even rounds is  $ \frac{1}{2}(\epsilon + \frac{1-\epsilon}{2}) =
\frac{1+\epsilon}{4} $ and the best that Eve can do is to
minimize this average to 25 percent by minimizing $\epsilon.$

 \end{document}